# Physical Insight into the 'Growing' Evanescent Fields of Double-Negative Metamaterial Lenses Using their Circuit Equivalence

Andrea Alù, *Student Member, IEEE,* Nader Engheta, *Fellow, IEEE*

*Abstract*— Pendry in his paper [Phys. Rev. Lett., 85, 3966 (2000)] put forward an idea for a lens made of a lossless metamaterial slab with $n = -1$, that may provide focusing with resolution beyond the conventional limit. In his analysis, the evanescent wave inside such a lossless double-negative (DNG) slab is "growing", and thus it "compensates" the decaying exponential outside of it, providing the sub-wavelength lensing properties of this system. Here, we examine this debated issue of "growing exponential" from an equivalent circuit viewpoint by analyzing a set of distributed-circuit elements representing evanescent wave interaction with a lossless slab of DNG medium. Our analysis shows that, under certain conditions, the current in series elements and the voltage at the element nodes may attain the dominant increasing due to the suitable resonance of the lossless circuit, providing an alternative physical explanation for "growing exponential" in Pendry's lens and similar sub-wavelength imaging systems.

## I. Introduction

THE idea of left-handed (LH) media, which dates back to 1967 when Veselago [1] theoretically studied plane wave propagation in materials in which he assumed both permittivity and permeability simultaneously having negative real parts, has attracted a great deal of attention in recent years. Various problems and ideas involving such media have been proposed and studied by many research groups. One such idea, namely a lens with possibility of perfect focusing, was theoretically suggested by Pendry in [2]. In his analysis, Pendry shows how evanescent waves, which are effectively responsible for sub-wavelength resolution, impinging on a suitably designed slab of double-negative (DNG) [3] material, may grow exponentially inside such a slab, and how this effect may "compensate" the decaying exponential taking place outside the slab [2]. This issue of "growing exponential" and subwavelength imaging has become the subject of interest for several research groups working in metamaterial research (see, e.g., [4]-[7]). Analogous sub-wavelength focusing and growing evanescent distributions have been demonstrated in two-dimensional negative-refractive-index transmission line structures [8]-[9].

In one of our previous works, we have shown how a similar phenomenon of "growing exponential" may occur in pairs of "conjugate" metamaterial slabs, i.e., pairs of DNG and double-positive (DPS) slabs or pairs of single-negative (SNG) layers such as epsilon-negative (ENG) and mu-negative (MNG) layers [10]. In these cases, we have shown how wave tunneling, transparency, and virtual image sub-wavelength displacement may be achieved under a proper choice of combinations of metamaterial parameters and slab thicknesses, independent on the property of the "outside" medium that surrounds the pairs of slabs. We attributed these findings to the presence of an "interface resonance" at the boundary between the two conjugate slabs. It is worth noting that these pairs of conjugate slabs also supported growing evanescent fields internally when an incident wave impinges on them. We showed how the "interface resonance" may be explained using the circuit-element analogy, when the conjugate distributed-circuit elements are paired to produce similar resonances and growing distributions for the voltage and currents.

In the present work, we explain how Pendry's lens may be viewed as a special case of a more general analysis of pairs of conjugate slabs described in our previous work [10] and how the growing evanescent field behavior in his lens may be clearly understood using anthe equivalent circuit analogy. Some of our preliminary results in this work were presented in a recent symposium [11].

## II. Formulation of the Problem

As in Pendry's lens case, here we consider a transverse

Manuscript received xx yy, 2004. This work is supported in part by the Fields and Waves Laboratory, Department of Electrical and Systems Engineering, University of Pennsylvania. A. Alù was supported by the scholarship "Isabella Sassi Bonadonna" from the Italian Electrical Association (AEI).

A. Alù is currently with the Department of Electrical Engineering, University of Roma Tre, Rome, Italy (e-mail: alu@uniroma3.it) and the Department of Electrical and Systems Engineering, University of Pennsylvania, Philadelphia, Pennsylvania 19104, U.S.A. (e-mail: andreaal@ee.upenn.edu).

N. Engheta is with the Department of Electrical and Systems Engineering, University of Pennsylvania, Philadelphia, Pennsylvania, 19104, U.S.A. (corresponding author, e-mail: engheta@ee.upenn.edu).

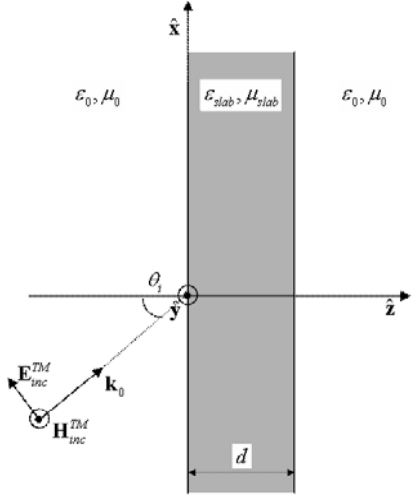

Fig. 1. Geometry of the problem: a plane wave impinging from vacuum on a planar slab with $\varepsilon_{slab}$, $\mu_{slab}$. In Pendry's "perfect" lens $\varepsilon_{slab} = -\varepsilon_0$, $\mu_{slab} = -\mu_0$ at the frequency of interest.

magnetic (TM) plane wave impinging on a metamaterial slab, as shown in Fig. 1. The incident electric and magnetic fields can be written as:

$$\boldsymbol{H}_{inc}^{TM} = \hat{\boldsymbol{y}} H_0 e^{-jk_x x - \sqrt{k_x^2 - k_0^2}\, z}$$
$$\boldsymbol{E}_{inc}^{TM} = \left( \hat{\boldsymbol{x}} \frac{\sqrt{k_x^2 - k_0^2}}{j\omega\varepsilon_0} - \hat{\boldsymbol{z}} \frac{k_x}{\omega\varepsilon_0} \right) H_0 e^{-jk_x x - \sqrt{k_x^2 - k_0^2}\, z}, \quad (1)$$

where the time harmonic $e^{j\omega t}$ is assumed with frequency $f = \omega/2\pi$, and $\varepsilon_0$, $\mu_0$, $k_0 = \omega\sqrt{\varepsilon_0 \mu_0}$ are the vacuum permittivity, permeability and wave number, respectively. Without loss of generality, we assume that the problem is two-dimensional, i.e., all quantities are independent of the $y$ coordinate. The transverse wave number $k_x$ may be smaller than $k_0$, for which the plane wave is propagating in the vacuum surrounding the slab, or greater than $k_0$, in which case the incident wave is evanescent. Eq. (1) is written for an evanescent incident wave, but remains valid when $k_x < k_0$ choosing the positive sign for the square roots, to ensure the radiation condition for the incident propagating plane wave. The total magnetic field in all the three regions of Fig. 1 may be expressed as:

$$\boldsymbol{H}_{z<0}^{TM} = \hat{\boldsymbol{y}} H_0 e^{+ik_x x} \left( e^{-\sqrt{k_x^2 - k_0^2}\, z} - R^{TM} e^{\sqrt{k_x^2 - k_0^2}\, z} \right)$$
$$\boldsymbol{H}_{slab}^{TM} = \hat{\boldsymbol{y}} H_0 e^{+ik_x x} \left( C_+^{TM} e^{-\sqrt{k_x^2 - k_{slab}^2}\, z} + C_-^{TM} e^{\sqrt{k_x^2 - k_{slab}^2}\, z} \right), \quad (2)$$
$$\boldsymbol{H}_{z>0}^{TM} = \hat{\boldsymbol{y}} T^{TM} H_0 e^{+ik_x x} e^{-\sqrt{k_x^2 - k_0^2}\,(z-2d)}$$

where $k_{slab}^2 = \omega^2 \varepsilon_{slab} \mu_{slab}$. We note that when evanescent waves are considered inside the slab, with $k_x^2 > k_{slab}^2$, the choice of sign for the square root $\sqrt{k_x^2 - k_{slab}^2}$ is not important, since both the forward-decaying and back-ward decaying waves are considered here. In the vacuum region to the right of the slab, the presence of only the transmitted wave with decaying exponential is assumed.

By satisfying all the boundary conditions at the interfaces $z = 0$ and $z = d$, one can obtain the values for coefficients $R^{TM}$, $C_\pm^{TM}$ and $T^{TM}$. Similar steps may be easily taken for the transverse electric (TE) polarization. For Pendry's lens, where $\varepsilon_{slab}/\varepsilon_0 = \mu_{slab}/\mu_0 = -1$, these coefficients have the following values:

$$\begin{aligned} R^{TE} = R^{TM} = 0, &\quad T^{TE} = T^{TM} = 1 \\ C_+^{TE} = C_+^{TM} = 0, &\quad C_-^{TE} = C_-^{TM} = 1 \end{aligned} \quad \forall k_x, \quad (3)$$

which, as described by Pendry, implies that on the plane $z = 2d$ each plane wave, whether propagating or evanescent, would have exactly the same value as the one it has at $z = 0$, essentially showing how such a "matched" DNG slab acts as a "perfect" lens, providing an image of the object plane at $z = 0$ without any limit on the resolution. Because $C_+^{TE} = C_+^{TM} = 0$ and $C_-^{TE} = C_-^{TM} = 1$, the evanescent wave inside this DNG slab is "growing" for both polarizations. As we show in the following, an equivalent circuit representation may provide a similar, but arguably more familiar, behavior for voltages and currents in suitably selected distributed circuit elements, thus providing new physical insights into this phenomenon.

### III. CIRCUIT EQUIVALENCE

For the TM plane wave propagation in a homogeneous isotropic medium, from Maxwell's equations one can write:

$$\frac{\partial E_x}{\partial z} = -j\omega\tilde{\mu}_{eq} H_y, \quad \frac{\partial H_y}{\partial z} = -j\omega\tilde{\varepsilon}_{eq} E_x, \quad (4)$$

where $\tilde{\mu}_{eq}$ and $\tilde{\varepsilon}_{eq}$ are shorthands for $\tilde{\mu}_{eq} \equiv \mu\left(1 - \frac{k_x^2}{\omega^2 \mu\varepsilon}\right)$ and $\tilde{\varepsilon}_{eq} \equiv \varepsilon$, and $\varepsilon$, $\mu$ are the material permittivity and permeability [12]. (By duality, one can easily write the corresponding terms for the TE case as well, which is $\tilde{\mu}_{eq} \equiv \mu$ and $\tilde{\varepsilon}_{eq} \equiv \varepsilon\left(1 - \frac{k_x^2}{\omega^2 \mu\varepsilon}\right)$.) These expressions may be viewed as formally analogous to the transmission line equations $\partial V / \partial z = -j\omega L_{eq} I$, $\partial I / \partial z = -j\omega C_{eq} V$ with the equivalent series inductance per unit length $L_{eq}$ and equivalent shunt capacitance per unit length $C_{eq}$ being proportional to $\tilde{\mu}_{eq}$ and $\tilde{\varepsilon}_{eq}$, as follows:

TABLE I
EFFECTIVE TL MODELS IN LOSSLESS DPS, DNG, ENG, MNG SLABS FOR THE TE AND TM PROPAGATING AS WELL AS EVANESCENT WAVES

| | **DPS** ($\mu > 0, \varepsilon > 0$) | **DNG** ($\mu < 0, \varepsilon < 0$) | **ENG** ($\mu > 0, \varepsilon < 0$) | **MNG** ($\mu < 0, \varepsilon > 0$) |
|---|---|---|---|---|
| TE Propagating $k_x^2 < \omega^2 \mu\varepsilon$ | $L_{eq} > 0 \quad \kappa \in \Re$ <br> $C_{eq} > 0 \quad Z_t \in \Re$ <br> 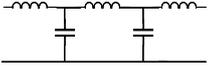 <br> L-C TL | $L_{eq} < 0 \quad \kappa \in \Re$ <br> $C_{eq} < 0 \quad Z_t \in \Re$ <br> 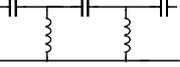 <br> C-L TL | Not applicable, since for $k_x \in \Re$, we always have $k_x^2 > \omega^2 \mu\varepsilon$ in lossless ENG | Not applicable, since for $k_x \in \Re$, we always have $k_x^2 > \omega^2 \mu\varepsilon$ in lossless MNG |
| TE Evanescent $k_x^2 > \omega^2 \mu\varepsilon$ | $L_{eq} > 0 \quad \kappa \in \Im$ <br> $C_{eq} < 0 \quad Z_t \in \Im$ <br> 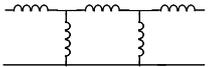 <br> L-L TL | $L_{eq} < 0 \quad \kappa \in \Im$ <br> $C_{eq} > 0 \quad Z_t \in \Im$ <br> 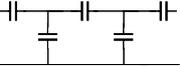 <br> C-C TL | $L_{eq} < 0 \quad \kappa \in \Im$ <br> $C_{eq} > 0 \quad Z_t \in \Im$ <br> 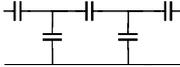 <br> C-C TL | $L_{eq} > 0 \quad \kappa \in \Im$ <br> $C_{eq} < 0 \quad Z_t \in \Im$ <br> 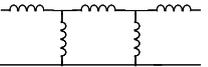 <br> L-L TL |
| TM Propagating $k_x^2 < \omega^2 \mu\varepsilon$ | $L_{eq} > 0 \quad \kappa \in \Re$ <br> $C_{eq} > 0 \quad Z_t \in \Re$ <br> 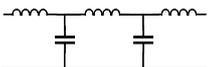 <br> L-C TL | $L_{eq} < 0 \quad \kappa \in \Re$ <br> $C_{eq} < 0 \quad Z_t \in \Re$ <br> 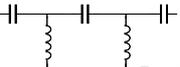 <br> C-L TL | Not applicable, since for $k_x \in \Re$, we always have $k_x^2 > \omega^2 \mu\varepsilon$ in lossless ENG | Not applicable, since for $k_x \in \Re$, we always have $k_x^2 > \omega^2 \mu\varepsilon$ in lossless MNG |
| TM Evanescent $k_x^2 > \omega^2 \mu\varepsilon$ | $L_{eq} < 0 \quad \kappa \in \Im$ <br> $C_{eq} > 0 \quad Z_t \in \Im$ <br> 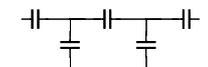 <br> C-C TL | $L_{eq} > 0 \quad \kappa \in \Im$ <br> $C_{eq} < 0 \quad Z_t \in \Im$ <br> 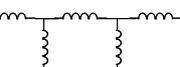 <br> L-L TL | $L_{eq} > 0 \quad \kappa \in \Im$ <br> $C_{eq} < 0 \quad Z_t \in \Im$ <br> 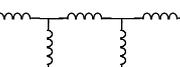 <br> L-L TL | $L_{eq} < 0 \quad \kappa \in \Im$ <br> $C_{eq} > 0 \quad Z_t \in \Im$ <br> 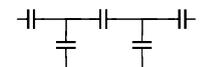 <br> C-C TL |

$$L_{eq} \propto \tilde{\mu}_{eq}, \quad C_{eq} \propto \tilde{\varepsilon}_{eq}. \tag{5}$$

It is worth noting that the transmission-line analogy may in general offer an interesting physical interpretation and alternative insight, effectively linking the voltage and current distributions along a circuit network to their local counterparts represented by the electric and magnetic fields. This is well known in the DPS case [12], but it is easily extended to the metamaterial parameters, as it has been shown in [8]-[10].

We note that even in a conventional DPS material, where $\mu$ and $\varepsilon$ are positive, the value of $L_{eq}$ in the TM case and $C_{eq}$ in the TE case may become negative, when $k_x^2 > \omega^2 \mu\varepsilon$, i.e., for an evanescent wave. As is well known, a negative equivalent inductance or capacitance at a given frequency may be interpreted effectively as a positive (dispersive) capacitance or inductance at that frequency, respectively [8], [10]. Therefore, for the TM case the evanescent plane wave propagation in a DPS medium may be modeled using a transmission line with a negative series inductance per unit length and a positive shunt capacitance per unit length, which effectively implies a positive series capacitance per unit length and a positive shunt capacitance per unit length. In such a C-C line, which is a ladder network made of capacitors, currents and voltages cannot "propagate" along the line, but instead they have an evanescent behavior, consistently with the electromagnetic counterpart. When a DNG material or an ENG or MNG medium is used, their suitable equivalent TL models may exhibit anomalous properties consistent with the features of wave propagation in such media. In general one may consider Table I showing the equivalent TL model for plane waves in lossless homogeneous isotropic media, with all possibilities for signs of the real part of their permittivity and permeability, both for the cases of propagating and evanescent waves. When losses are present, $\mu$ and/or $\varepsilon$ have complex values, which translates into positive series resistance and/or shunt conductance in the TL model.

If we now consider Pendry's lens problem, the equivalent 1-D TL model may be depicted in Fig. 2, where a TM evanescent wave, impinging on a "matched" DNG slab, is considered. [The TE case may be again obtained by duality, i.e., every inductor (capacitor) should be replaced by a capacitor (inductor).] In the figure, we have considered $k_x^2 > \omega^2 \mu_0 \varepsilon_0 = \omega^2 \mu_{slab} \varepsilon_{slab}$, which gives an evanescent wave in the vacuum and inside the slab. The primary parameters of the TL sections may be derived from Eq. (5). When the equivalent inductors or capacitors are negative, in the figure they are respectively shown as effective capacitors or inductors, i.e., since $C_{eq_{slab}} < 0$ and $L_{eq_{vac}} < 0$, we have $\left(j\omega C_{eq_{slab}}\right)^{-1} = j\omega L_{eff_{slab}}$, and $j\omega L_{eq_{vac}} = \left(j\omega C_{eff_{vac}}\right)^{-1}$. Moreover, since for Pendry's "matched" DNG slab $C_{eq_{slab}} = -C_{eq_{vac}}$ and $L_{eq_{vac}} = -L_{eq_{slab}}$ for any $k_x$, their values satisfy the following relations:

$$L_{eq_{slab}} C_{eff_{vac}} = L_{eff_{slab}} C_{eq_{vac}} = \omega^{-2} \qquad \forall k_x. \qquad (6)$$

From this relation, we get the following expressions for the secondary parameters $Z$ and $\kappa$ of each line segment:

$$\begin{cases} |Z|_{slab} \equiv \omega\sqrt{L_{eq_{slab}} L_{eff_{slab}}} = \left(\omega\sqrt{C_{eq_{vac}} C_{eff_{vac}}}\right)^{-1} \equiv |Z|_{vac} \\ \kappa^2_{slab} \equiv -L_{eq_{slab}}/L_{eff_{slab}} = -C_{eq_{vac}}/C_{eff_{vac}} \equiv \kappa^2_{vac} \end{cases} \forall k_x, (7)$$

which ensure that the magnitudes of the characteristic impedances and of the wave numbers are the same in the two lines, and that they are all imaginary quantities (since the wave is evanescent). Nothing is said in Eq. (7) on their signs, but they may be derived from the following considerations.

The signs of the imaginary wave numbers $\kappa$ in the C-C and L-L ladders have to be negative, to ensure the exponential causal decay in an infinite or a matched line with $e^{-j\kappa z}$ propagating factor. For what concerns the signs of the characteristic impedances for an L-L or C-C line, we obviously expect to have inductive or capacitive characteristic impedance for these lines, respectively. (A more rigorous demonstration may be obtained by adding a small amount of losses to the TL parameters, in order to select the proper branches of the square roots with positive real part, similarly to what shown in [3].) This results in the following expressions (where we choose always the positive sign for the square roots):

$$\kappa_{vac} = -j\sqrt{C_{eq_{vac}}/C_{eff_{vac}}}, \; \kappa_{slab} = -j\sqrt{L_{eq_{slab}}/L_{eff_{slab}}},$$
$$Z_{vac} = \left(j\omega\sqrt{C_{eq_{vac}} C_{eff_{vac}}}\right)^{-1}, \; Z_{slab} = j\omega\sqrt{L_{eq_{slab}} L_{eff_{slab}}}. \qquad (8)$$

These formulas clearly show that, unlike the case of *propagating* wave interaction with this DNG slab where the impedances are matched [2], [3], for the *evanescent* waves the two media are not impedance-matched, since $Z_{vac} = -Z_{slab}$, but on the other hand at the interface a resonance arises, giving rise to a reflection coefficient $R = \dfrac{Z_{vac} - Z_{slab}}{Z_{vac} + Z_{slab}} = \infty$. This "interface" resonance is the key in understanding the anomalous behavior of this setup, and the circuit analogy gives a further insight into this phenomenon.

Eq. (6) and the previous consideration, in fact, imply that at each of the two interfaces between vacuum and the DNG slab, the adjacent series elements $C_{eff_{vac}}$ and $L_{eq_{slab}}$ would resonate at the frequency of operation for any $k_x$ (this is of course the circuital counterpart of the surface waves supported by such an interface [13], which indeed play a key role in the physics of the sub-wavelength imaging [5]-[7]). Therefore, looking at the right interface (node $n = n'$ in the figure), we note that the voltage at the left node of this $L_{eq_{slab}}$ ($n-1$ in the figure) is the same as the voltage at the right node of this $C_{eff_{vac}}$ ($n'-1$ in the figure). Consequently, the next two adjacent shunt elements $L_{eff_{slab}}$ and $C_{eq_{vac}}$ are now in parallel and they are also in resonance, again according to Eq. (6). Repeating this argument, we note that effectively a segment of the C-C line with length $d$ would be in resonance with the entire L-L line that represents the matched DNG slab with the same thickness $d$. In fact, we expect that the voltages and current at every node $i$ are the same as those at the corresponding node $i'$. Therefore, the voltage and current at one end (node 0) of this "resonant pair" of C-C and L-L lines (each with length $d$) would be the same at those at its other end (node 0'), which implies that this pair appears to become "transparent" to the rest of the structure. This also means that if in the C-C segment we have a decaying exponential voltage (which is the only physical possibility), we should have a "growing exponential" voltage in the L-L segment in order to have the voltage nodes the same at the beginning and at the end of this pair. In fact, due to the multiple reflections at the two interfaces, each with an "infinite" reflection coefficient, the "reflected growing" exponential builds up in the steady-state regime and totally dominates the impinging decaying exponential in the DNG slab by itself. It is important to underline here that the presence of the "growing" exponential in the L-L line is due to the "interface" resonance at the boundary between the C-C and L-L lines, and it is not just only due to the L-L line (i.e, the DNG slab) by itself (as confirmed also by Eq. (8)). In other words, in a dual scenario if we had a "vacuum" slab sandwiched between two semi-infinite DNG half spaces, following a similar argument we would have seen the growing exponential in the vacuum slab region!

A further confirmation of the presence of the growing exponential in the L-L line segment sandwiched as in Figure 2 may be found directly by solving such a circuit network. Let us excite this circuit with a steady-state time-harmonic voltage source $V_{exc}$ at a given node in the semi-infinite C-C transmission line on the left of the L-L segment. We have shown above that the pair of L-L segment together with the $d$-long C-C segment is in resonance and thus "transparent" to the rest of the structure. Therefore, we expect to have the voltage and current in the left C-C segment to be related as follows:

$$I_{exc} = \dfrac{V_{exc}}{Z_{vac}} = j\omega\sqrt{C_{eq_{vac}} C_{eff_{vac}}}\, V_{exc} = \dfrac{jV_{exc}}{\omega\sqrt{L_{eq_{slab}} L_{eff_{slab}}}}, \qquad (9)$$

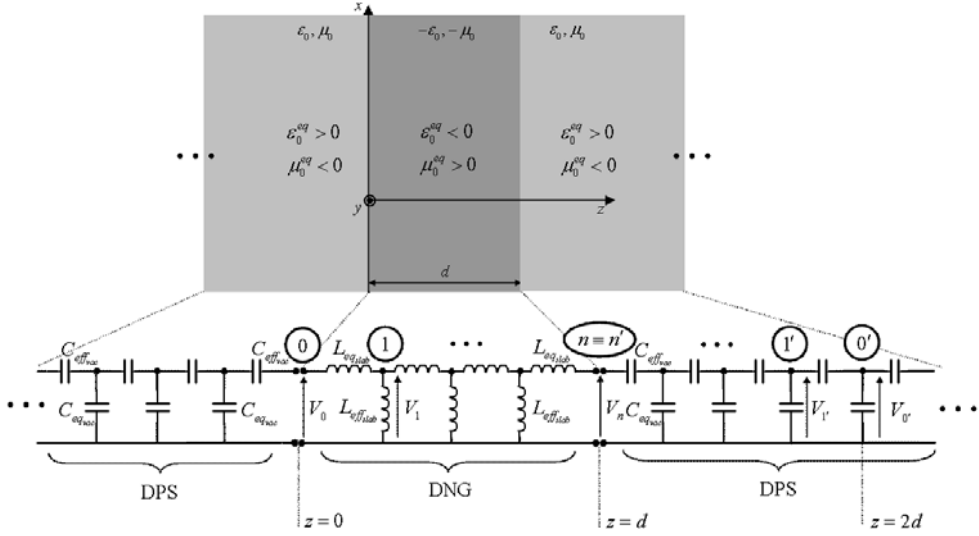

Fig. 2. Equivalent TL model for a TM evanescent wave impinging on the Pendry's perfect lens.

as in any matched or infinite line. For the same reason, the node voltages and branch currents along the C-C line in the left decay exponentially, due to the imaginary value of $\kappa_{vac}$ shown in Eq. (8), until we get to the left interface between the C-C and L-L lines, which is denoted in the figure with node "0". Let us denote the node voltage at this interface $V_0$, and the series branch current $I_0$, which can be expressed as $I_0 = j\omega\sqrt{C_{eq_{vac}} C_{eff_{vac}}} V_0$, following (9). As we move into the C-C line, we can evaluate the node voltage $V_n$ and the current $I_n$ in the following branch at the $n^{th}$ node with the recursive relations:

$$V(n) = V(n-1) - j\omega L_{eq_{slab}} I(n-1)$$
$$I(n) = I(n-1) - \frac{V(n)}{j\omega L_{eff_{slab}}}, \qquad (10)$$
$$V(0) = V_0$$
$$I(0) = j\omega\sqrt{C_{eq_{vac}} C_{eff_{vac}}} V_0 = \left(-j\omega\sqrt{L_{eq_{slab}} L_{eff_{slab}}}\right)^{-1} V_0 = I_0$$

This is analogous to the Fibonacci problem, and the explicit solution is written as:

$$V(n) = \frac{V_0}{2^{n+1} L_{eff_{slab}}^n} \cdot$$
$$\left[\left(1 + \frac{2\sqrt{L_{eff_{slab}}} - \sqrt{L_{eq_{slab}}}}{\sqrt{4L_{eff_{slab}} + L_{eq_{slab}}}}\right) a_+^n + \left(1 - \frac{2\sqrt{L_{eff_{slab}}} - \sqrt{L_{eq_{slab}}}}{\sqrt{L_{eq_{slab}} + 4L_{eff_{slab}}}}\right) a_-^n\right]$$
$$I(n) = \frac{I_0}{2^{n+1} L_{eff_{slab}}^n} \cdot$$
$$\left[\left(1 + \frac{2\sqrt{L_{eff_{slab}}} + \sqrt{L_{eq_{slab}}}}{\sqrt{4L_{eff_{slab}} + L_{eq_{slab}}}}\right) a_+^n + \left(1 - \frac{2\sqrt{L_{eff_{slab}}} - \sqrt{L_{eq_{slab}}}}{\sqrt{4L_{eff_{slab}} + L_{eq_{slab}}}}\right) a_-^n\right] \quad (11)$$

$$a_\pm = 2L_{eff_{slab}} + L_{eq_{slab}} \pm \sqrt{L_{eq_{slab}}\left(4L_{eff_{slab}} + L_{eq_{slab}}\right)}.$$

These expressions may be compacted after some algebraic manipulations, becoming:

$$V(n) = \frac{V_0(4\gamma+1)^{n/2-1}}{(2\gamma)^{n-1}} \cdot$$
$$\sum_{k=0 \text{ even}}^{n-2} \binom{n-1}{k}\left[\frac{n-k-1}{k+1}(2\gamma+1) + 2\sqrt{\gamma} - 1\right]\left(\frac{2\gamma+1}{\sqrt{4\gamma+1}}\right)^k$$
$$I(n) = \frac{I_0(4\gamma+1)^{n/2-1}}{(2\gamma)^{n-1}} \cdot \qquad , \quad (12)$$
$$\sum_{k=0 \text{ even}}^{n-2} \binom{n-1}{k}\left[\frac{n-k-1}{k+1}(2\gamma+1) + 2\sqrt{\gamma} + 1\right]\left(\frac{2\gamma+1}{\sqrt{4\gamma+1}}\right)^k$$

where $\gamma = L_{eff_{slab}}/L_{eq_{slab}}$ and these formulas are valid for even $n$. The values of $V(n)$ and $I(n)$, as in any pseudo-Fibonacci series, grows exponentially with $n$. This growth continues until we reach the right interface of the L-L segment. Beyond this interface, we are in the C-C segment, and with a similar argument we expect to have a decaying exponential with symmetrical values as in the L-L segment, i.e., $V(j') = V(j)$, $I(j') = I(j)$ for any $0 \le j \le n$. Therefore, the maximum values of $V(n)$ and $I(n)$ are expected to be at the interface $n = n'$, as predicted by Pendry in his DNG slab [2]. We have thus far shown how a "growing" exponential behavior inside a matched DNG slab of thickness $d$ may be justified using the circuit equivalence with the L-L and C-C lines. In fact, as we have mentioned in our previous work, one can suggest that this field behavior may also exist when a DNG slab is juxtaposed with a "conjugate" DPS slab of the same thickness (in the present case, this DPS layer is part of the outside vacuum region), or an epsilon-negative (ENG) slab paired with a mu-negative (MNG) slab of equal thickness [10] (and these cases would show total transparency independently of the parameters of the outside region surrounding the system).

In principle, this anomalous transparency is independent of the thickness of the two slabs (or in Pendry's lens of the DNG slab itself), as long as the slabs have equal thickness $d$. However, an important question may arise here: May we still have a growing exponential behavior inside the DNG slab (or equivalently inside the L-L line here), if its thickness becomes infinite, i.e., if we have a semi-infinite DNG space? According to the analysis presented here, the answer is as follows: we need both interfaces to achieve this exponential growth, therefore this effect may not take place if the second interface is at infinity. Moreover, due to the multiple resonances/reflections that are necessary for the phenomenon to build up, a thicker slab should be more sensitive to the inherent losses of the setup and more time should be also required for the phenomenon to build up and reach the steady-state regime. Therefore, even if the second interface is too far apart (and not at infinity) in practical systems the growing exponential may disappear. This is consistent with the findings we have shown in [14]

We know, however, that an interface between the semi-infinite matched DNG and DPS media may indeed support a surface plasmon wave [13]. In this case, for an incoming *evanescent* wave, the transverse impedances of the two regions are complex conjugate of each other, i.e., $Z_{L-L} \equiv -Z_{C-C} = jX$, and therefore the Fresnel "reflection" and "transmission" coefficients for such an incident evanescent wave become infinite, as we have previously found for each of the two interfaces in the circuit analog. We reiterate that this in principle does not violate any physical law, since these coefficients here describe the relation between an "incident" *evanescent* wave and its "reflected" and "transmitted" *evanescent* waves, neither of which carries any real power. So when we have a source in front of the interface between two semi-infinite matched DNG and DPS media, the resonant surface wave may be excited along the interface, resulting in an infinitely large field value. However, on both sides of this interface, the fields, albeit infinitely large, *decay* exponentially, since the field distribution represents a surface wave propagating along such an interface.

## IV. Effects of Loss and Mismatch in Materials

Thus far we have assumed complete losslessness and match between the DNG and the outside region. When loss and/or mismatch in the material parameters is present, we expect to have certain variations to the field distribution in this geometry, and as the DNG slab gets thicker, such variations would be more sensitive to the presence of loss and mismatch, as also anticipated. This sensitivity is mainly due to the presence of the surface wave supported by the slab (notice that the matched slab without losses, in fact, does not support any surface wave, even thought the two interfaces delimiting the slab would do so). If the structure supports a surface wave with a given $k_x > k_0$, in fact, the reflection and transmission coefficients for the DNG slab would no longer be flat for all $k_x$, but instead would experience a peak (or a singularity in case of no loss) at the value of $k_x$ for which the surface wave is supported. This has been shown by others in several recent papers studying this phenomenon [15]-[17] and indeed limits the overall resolution to certain extent, which can still be sub-wavelength value better than the conventional resolution.

The effects of loss and mismatch may again be explained by the equivalent circuit models described here. In the ideal lossless matched case, we showed that for any value of $k_x$, the $C_{eff_{vac}}$ series capacitors are all in resonance with the corresponding series $L_{eq_{slab}}$ and similarly every $C_{eq_{vac}}$ shunt capacitor is in resonance with a corresponding shunt $L_{eff_{slab}}$. The quality factor $Q$, of such a resonance is thus infinitely large. However, the loss in the system causes the quality factor $Q$ to become finite, resulting in a quicker drop of the transmission for high $k_x$ (for which the equivalent electrical length of the TL increases). Moreover, the mismatch does not allow a "perfect" resonance between the inductors and capacitors mentioned above for all value of $k_x$: only for certain $k_x$ such a resonance may still occur.

## V. Conclusions

Considering the transverse magnetic (TM) plane wave interaction with Pendry's "perfect" lens, we have shown how this problem may be treated equivalently as a finite segment of L-L line, representing the DNG slab for the evanescent wave, sandwiched between two semi-infinite segments of C-C lines, representing the outside DPS regions for the TM evanescent wave. In this analogy, voltages and currents represent the electric and magnetic fields. We have analyzed the overall circuit, showing the possibility of explaining the growing exponential term for the electromagnetic field along the DNG segment as a resonant phenomenon in the circuit, with an analogous growth of voltage and current distributions. The model is effective also in presence of losses, which are represented by resistances and conductances and may give further insights into the anomalous phenomenon of sub-wavelength imaging utilizing metamaterials.

## References

[1] V. G. Veselago, "The electrodynamics of substances with simultaneously negative values of ε and μ," *Soviet Physics Uspekhi*, vol. 10, no. 4, pp. 509-514, 1968. [in Russian: *Usp. Fiz. Nauk*, vol. 92, pp. 517-526, 1967.]